# Topological engineered 3D printing of Architecturally Interlocked Petal-Schwarzites

Rushikesh S. Ambekar[a], Leonardo V. Bastos[b], Douglas S. Galvao[c,d*], Chandra S. Tiwary[a*] and Cristiano F. Woellner[b*]

[a]Department of Metallurgical and Materials Engineering, Indian Institute of Technology Kharagpur-721302, West Bengal, India

[b]Physics Department, Federal University of Parana, Curitiba-PR, 81531-980, Brazil

[c]Applied Physics Department, Gleb Wataghin Institute of Physics, State University of Campinas, Campinas,SP, 13083-970, Brazil

[d]Center for Computing in Engineering & Sciences, State University of Campinas, Campinas, SP, 13083-970, Brazil

**Abstract**

The topologically engineered complex Schwarzites architecture has been used to build novel and unique structural components with a high specific strength. The mechanical properties of these building blocks can be further tuned, reinforcing with stronger and high surface area architecture. In the current work, we have built six different Schwarzites structures with multiple interlocked layers, which we named architecturally interlocked petal-schwarzites (AIPS). These complex structures are 3D printed into macroscopic dimensions and compressed using uniaxial compression. The experimental results show a strong dependency of mechanical response on the number of layers and topology of the layers. Fully atomistic molecular dynamics compressive simulations were also carried out, and the results are in good agreement with experimental observations. They can explain the underlying AIPS mechanism of high specific strength and energy absorption. The proposed approach opens a new perspective on developing new 3D-printed materials with tunable and enhanced mechanical properties.

Keywords: 3D printing, Interlocked Architecture, Schwarzites, Mechanical properties, MD simulation

## 1. Introduction

The biomimetic approach is one of the best ways to design lightweight structures that optimize physical and chemical properties essential for various engineering applications. Bio-inspired circular tubes (e.g., nacre based [1], wood-inspired [2] etc.), multi-cell tubes (e.g., bamboo [3], horsetail [4], etc.), cellular structure (e.g., honeycomb [5], turtle inspired foams [6]), and living species built architectures (i.e., ant mounds, seashells, nacre, honeycomb, spider web, palm, bone, muscles, DNA, etc.) [7]. Although all of these architectures are very different from each other, there are still a few common aspects among them, such as complex topology, controlled porosity, optimum interconnects, multi-layered structures, and best possible ways to distribute stress and strain.

Recently, taking inspiration from nature, researchers started building complex carbon-based structures, such as schwarzites [8], tubulanes [9], interconnected nanotubes [10], and many more such architectures. All these complex structures show high specific strength with high porosity, high surface-to-volume ratio, strong interconnects, and directional stress distributions. A systematic mechanics study of these complex structures highlights unique mechanical deformations [11]. Using the current state-of-the-art simulation packages, such as atomistic molecular dynamics, Monte Carlo, AI-ML-based package, etc., researchers have established the relation between structure, topology, and mechanical deformation/properties of such complex architectures. Recently, it has been shown that some of these properties are scale independent [12]. Thus it can be easily used to build structures at different length scales, such as micro, meso, and macro.

With the invention of 3D printing, researchers can now build/make very complex architectures, built or predicted by theoretical simulations. Schwarzites are porous crystalline 3D structures of carbon, proposed by Mackay and Terrones using mapping/placing carbon rings containing six, seven, or eight atoms into triply periodic minimal surfaces (TPMS) [13,14].

Recently, several groups, including ours, have used 3D printing to fabricate theoretically designed architectures at different length scales. These 3D-printed architectures were tested under different loading conditions to determine the deformation behavior of such complex architectures. The summary of several research works in this direction concludes that the stress distribution or strain accumulation is highly dependent on the topology, interconnect, and/or interlayer interactions [15]. Engineering these parameters can result in enhanced mechanical properties, such as high impact resistance [16], high energy absorption [17], negative Poisson's ratio [18], superelasticity [19], self-healing [20], superplastic, and many more.

In this work, we have used our theoretical and experimental knowledge to build a unique structure named Architecturally Interlocked Petal-Schwarzites (AIPS). We have used the specific region/blocks/interconnect (as shown in **Fig. 1**) from the primitive schwarzites family to build the current new architectures. We have modified these building blocks by tuning the hexagon-octagon ratio to engineer the topology of the structure shown in **Fig. 1(b-i)**. We have chosen these modified building blocks and placed them into an interlocked layered architecture (shown in **Figs. 2** and **3**), which is inspired by rose petals. Therefore we named them petal-schwarzites. Depending on the combination of the individual building blocks and the number of layers (2, 3, and 4), we were able to build six different architectures of Petal-Schwarzites, as shown in **Figs. 2** and **3**. In order to compare the mechanical deformation of these AIPS, we have made them into a periodic architecture of similar dimensions/scale, as shown in **Fig. 4**.

These proposed new architectures were 3D printed into macroscopic dimensions using fused filament fabrication-based printers (shown in **Fig. 4**). We have carried out compression tests to evaluate the mechanical performance of the 3D-printed structures. We have also compared the experimental (3D-printed) and molecular dynamics simulated compression (atomic models) behavior of the structures. We have established a relationship between stress distribution or strain accumulation with the topology, interconnect, and/or interlayer interactions using the experimental and theoretical results. We have also evaluated other AIPS mechanical properties, such as high yield strength, elastic modulus, and energy absorption.

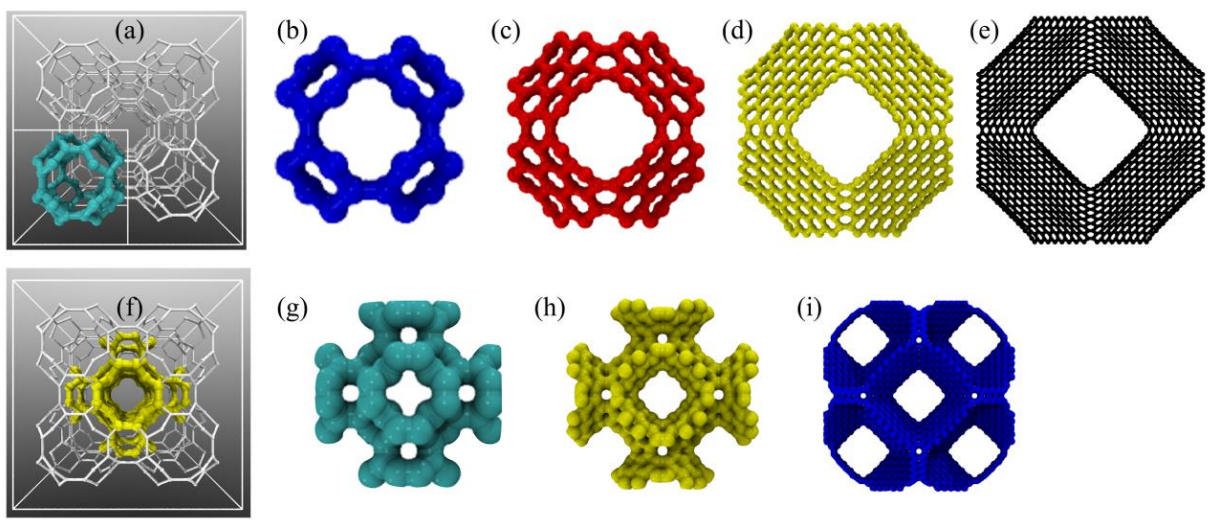

**Figure 1.** Schematic depicting the different building blocks used for building the Petal-Schwarzites. (a) The unit cell of schwarzites with a solid color block chosen for Petal-Schwarzites. Different ratios of hexagon-octagon result in different building blocks (b) P8-0 (c) P8-1 (d) P8-3 (e) P8-7. (f) The unit cell of schwarzites with a solid color block interconnects chosen for interconnecting layers of Petal-Schwarzites. Different ratios of hexagon-octagon result in different building blocks (g) $P_0 8$ (h) $P_1 8$ (i) $P_3 8$.

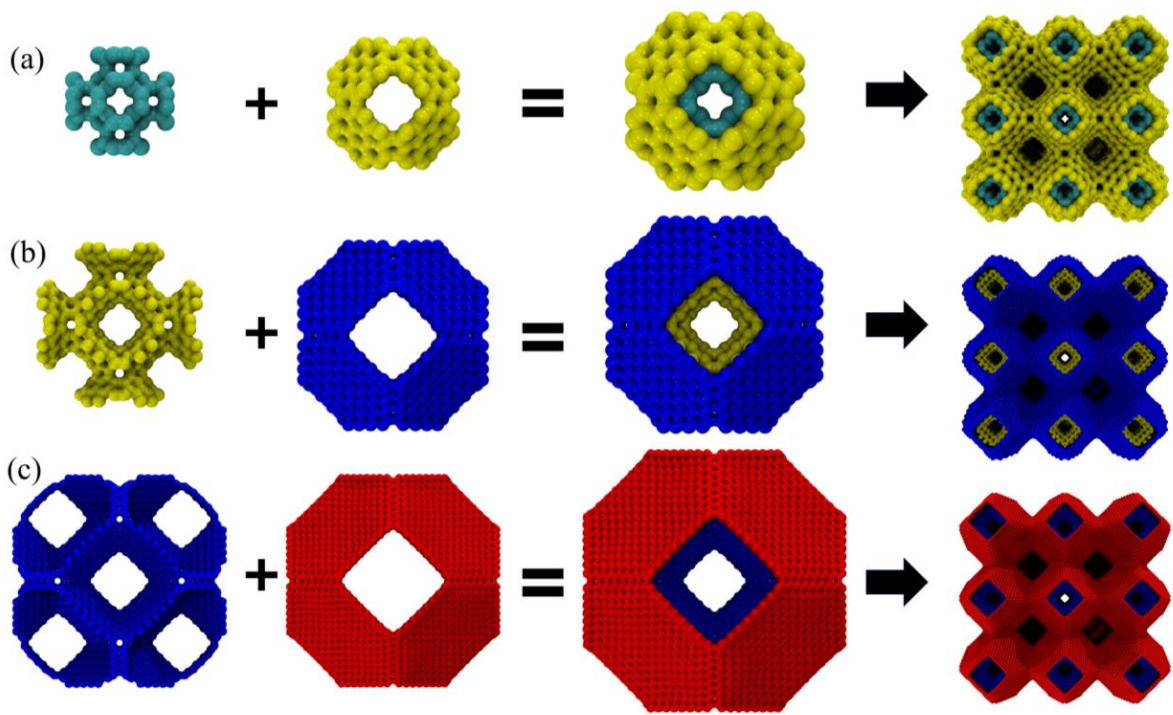

**Figure 2.** Schematic depicting the two layers used to build two interconnected layered Petal-Schwarzites. (a) $H_2P_08$ (b) $H_2P_18$ (b) $H_2P_38$.

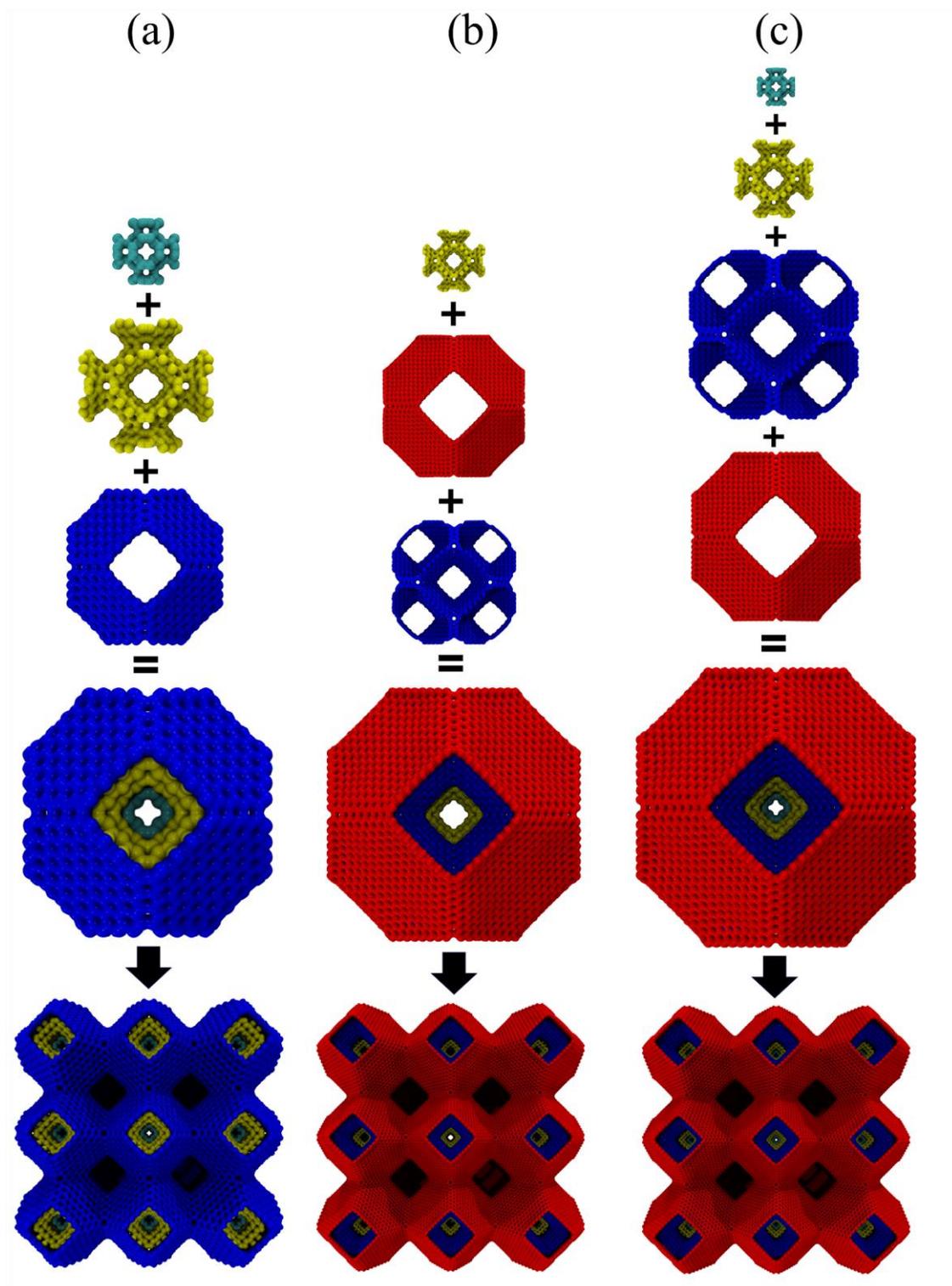

**Figure 3.** Schematic depicting the three and four layers used to build three and four interconnected layered Petal-Schwarzites. (a) $H_3P_08$ (b) $H_3P_18$ (c) $H_4P_08$.

## 2. Methodology
### 2.1 Experimental details

3D models of Petal-Schwarzites were generated using structural information from molecular dynamics (MD) simulations. The generated STL files were processed in Flashprint 4.1.0 software and converted into a G-code program to feed in the 3D printer. **Fig. 4(a)** shows the MD simulated and experimental structures. Fused deposition modeling was employed to fabricate the porous structures with the help of Flashforge adventure 3 3D printer. The solid polylactic acid filaments were used as an input material with a 1.75 mm standard diameter. The extruder and bed temperature was kept at 210°C and 50°C, respectively, and constant throughout the printing. Paper tape was used as a substrate to adhere to the 3D-printed structure. The porous structures were printed at high filling density (100%) and at high resolution (single layer thickness of 180 µm). The dimensions of all 3D-printed structures were 5 cm. The compression test was carried out at room temperature using UTM SHIMADZU (AG 5000G). The 3D printed structures were subjected to compressive loading under a constant rate of 1 mm/min. For detailed deformation analysis of the structures, photo snapshots were taken as a function of time.

### 2.2 Modeling details

The hybrid structures are composed of multiple schwarzites building blocks (top part). In order to create the interlocked Petal-Schwarzites (bottom), the original schwarzites' unit cells were modified in order to allow a perfect structural match with the next unit cell layers. Only the cases of P8-0 and P08 schwarzites are presented here. We performed MD uniaxial compressive tests at a constant rate with an NPT ensemble and using timesteps of 0.25 fs. Periodic boundary conditions were considered along all directions. Before compressing the structures, a thermalization process was performed for 5 ps. For all structures, it was tested whether the stress had reached a stationary and close to zero value by the end of the thermalization process and also whether their potential energy stabilized. After the thermalization process, the structures are then compressed along the z-direction at a constant strain rate of $10^{-2}$ ps$^{-1}$ for 100 ps. The structure's mechanical properties were studied during this part of the simulation by analyzing the stress-strain curve for each one of them.

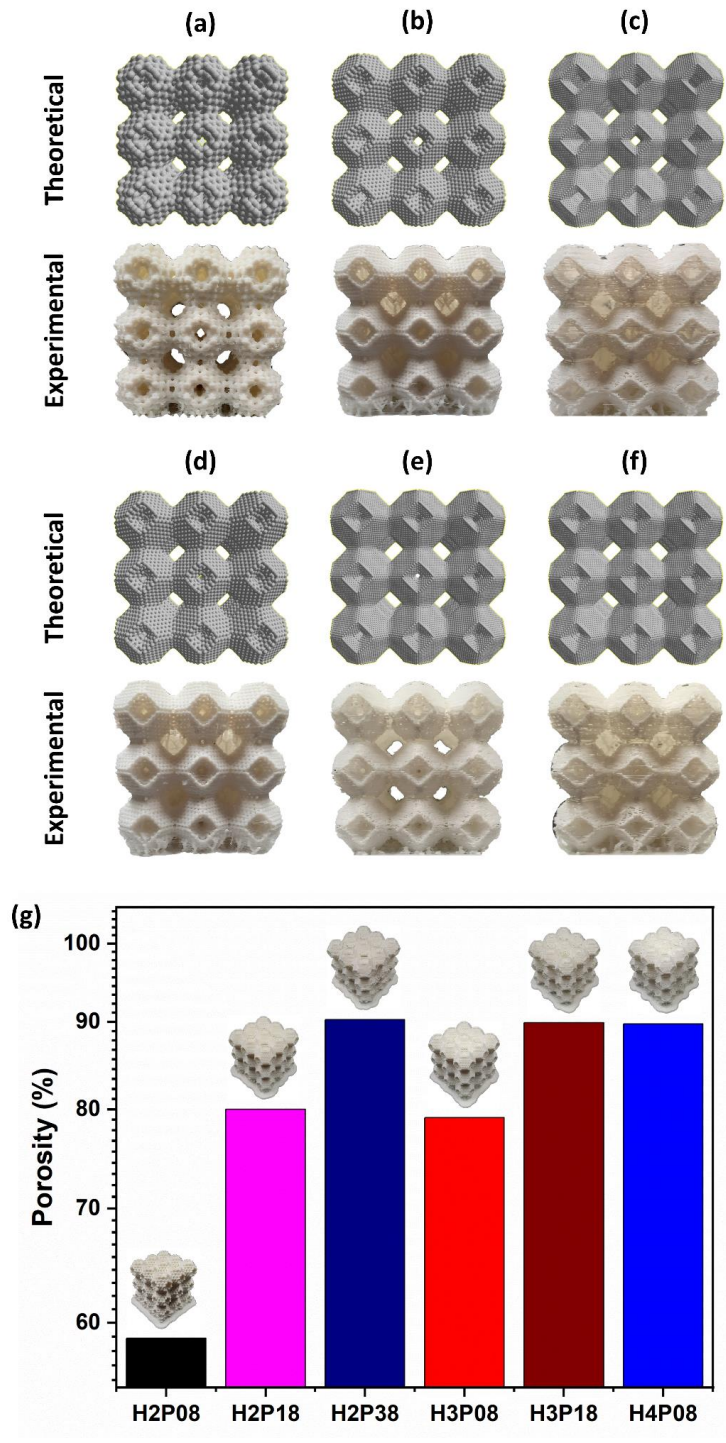

**Figure 4**. (a)-(f) Front view of the theoretical and experimental 3D printed interlocked Petal-Schwarzites structures (a-$H_2P_08$, b-$H_2P_18$, c-$H_2P_38$, d-$H_3P_08$, e-$H_3P_18$, f-$H_4P_08$); (g) Percentage porosity of the different Petal-Schwarzites structures.

## 3. Results and discussion

Bulk density was used to normalize the parameters like yield strength, toughness, and Young's modulus to compare the mechanical performance of lightweight structures. In **Fig. 5** we present the results for all six AIPS structures (left, experimental; right, MD results). The compressive behavior under uniaxial compression is shown in **Fig. 5(a)**. It is a semi-log plot, where the y-axis and x-axis represent stress and percentage strain, respectively. The behavior of the stress-strain curve for all the structures is the same up to 0.5% strain and values increase abruptly until 0.01 MPa. We observe a change in slope of the stress-strain curve for different structures, which qualitatively indicates change in modulus for different structures. After that the structure reaches yielding/yield strength followed by plastic plateau deformation/densification. We observe a drop in stress for $H_2P_18$ and $H_3P_08$ near 3% strain; apart from these two structures there is no significant drop observed for the rest of the structures up to 20%. After 5% strain stress remains more or less on the same level for all the structures.

     **Figure 5(b)-(d)** shows the mechanical properties (specific Young's modulus, specific yield strength, and specific energy absorption) of all the structures. Among all the structures, $H_3P_08$ has the highest, and $H_4P_08$ has the lowest specific Young's modulus value (See **Fig. 5(b)**). Specific yield strength of $H_3P_08 > H_2P_18 > H_2P_08$ is nearly 16% higher than $H_3P_18 > H_2P_38 > H_4P_08$, as shown in **Fig. 5(c)**. **Figure 5(d)** shows all the structures' specific energy absorption values. $H_2P_08$ has the highest value, whereas $H_2P_38$ has the lowest one. The deformation response of 3D printed AIPS structures are shown in Supplementary Videos **SV1**.

     In order to gain further insights into the AIPS mechanical response and deformation mechanisms, we have carried MD simulations for all six proposed structures. The stress-strain curves are shown in **Fig. 5(e)**. The trend of the theoretical stress-strain plot is similar to that of the

experimental one. The AIPS structures deform elastically up to less than 0.5% strain, then followed by yielding and reaching the densification region. The comparative specific Young's modulus, Young's modulus, and specific energy absorption (SEA) are shown in **Fig. 5(f)-(h)**.

We have normalized and compared specific Young's modulus, as shown in **Fig. 5(f)**. $H_2P_08$ and $H_2P_18$ show similar values compared to $H_2P_38$. The yield modulus for two layered AIPS shows the following trend $H_2P_08 > H_2P_18 > H_2P_38$, which is similar to their density variation (as displayed in **Fig. 5(g)**). The SEA shows an inverse order, i.e., $H_2P_08 < H_2P_18 < H_2P_38$. This can be explained by considering that higher porosity results in increased energy absorption (as illustrated in **Fig. 5(h)**).

Among two layered AIPS, $H_2P_18$ is the optimum structure, it has low density, good strength, and good energy absorption. The yield modulus for three layered AIPS shows the following trend $H_3P_08 > H_3P_18$, which is similar to their density variation. We have normalized and compared specific yield modulus as shown in **Fig. 5(f)**. The difference has decreased, but still $H_3P_08 > H_3P_18$. The SEA shows an inverse order, i.e., $H_3P_08 < H_3P_18$. This can be explained again by considering that higher porosity results in increased energy absorption. The variation for AIPS three layers is similar to the case of two layers, although three layers' architecture shows improved mechanical response (SEA and specific Young's modulus). The AIPS four-layered architecture ($H_4P_08$) shows a similar specific modulus to that of $H_3P_18$ and increased specific energy absorption.

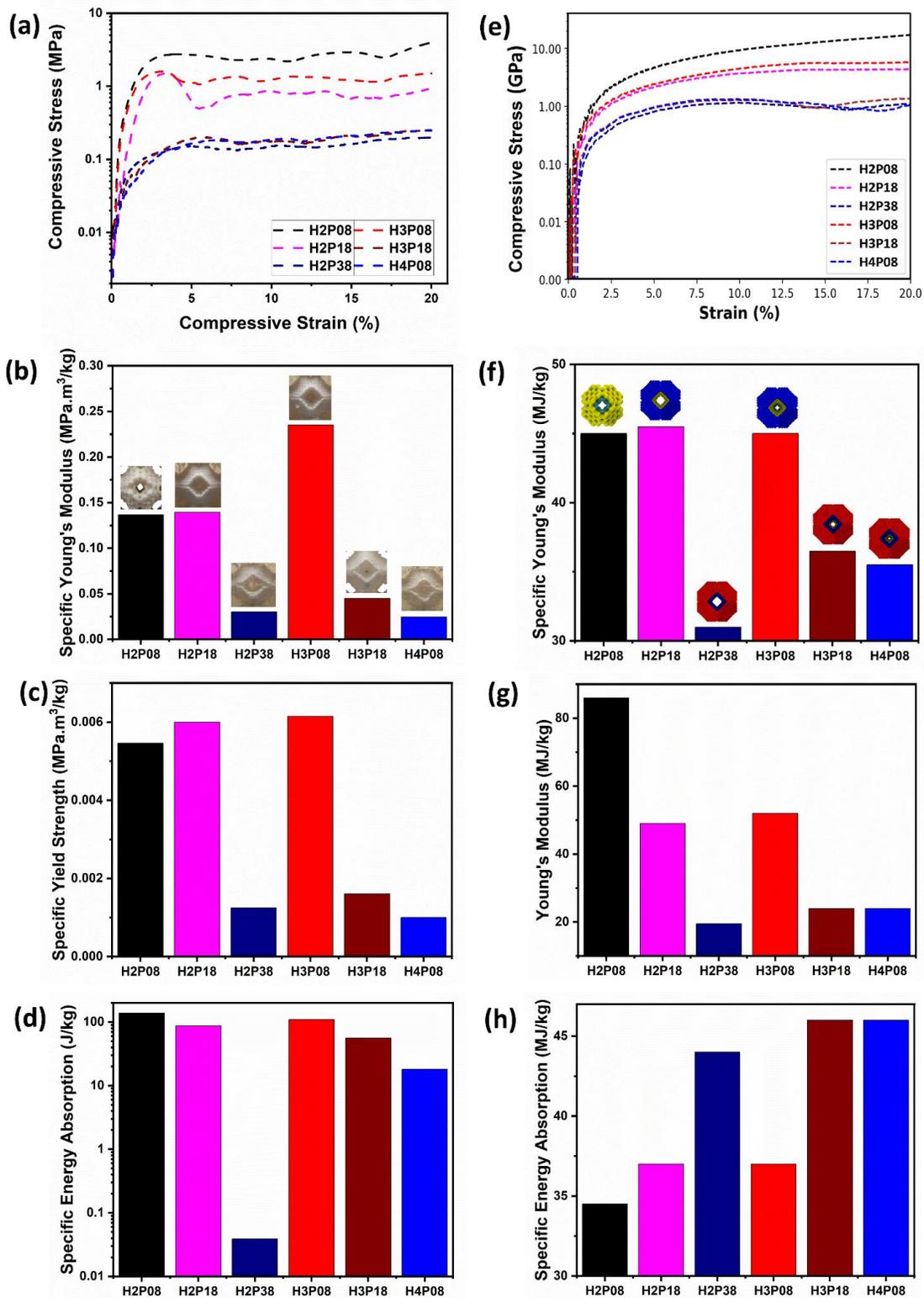

**Figure 5.** (a) Experimental compressive stress-strain curves of the AIPS structures. (b) Specific yield strength; (c) Specific Young's modulus; (d) Specific energy absorption; (e) AIPS stress-

strain curves from MD simulations. (f) Specific Young's modulus; (g) Young's modulus, and; (h) AIPS specific energy absorption from MD simulations.

During the experimental deformation tests, we have taken photo snapshots at various % compression strain regimes. For simplification, **Fig. 6(a)** shows different sections of the structures, such as upper, middle, and lower horizontal sections, vertical ones, and joining pillars. As shown in **Fig. 6(b),** $H_2P_08$ shows layer-by-layer deformation mechanisms. The first upper portion of the horizontal section deforms. After that, pore deformation occurs. Whereas, in the case of $H_3P_08$, the load is highly accumulated at the pillars joining two segments. Initially, the load is redistributed by the pillars. Hence it deforms, and in later stages, the layers in the middle section become fractured. Initially, a uniform deformation of $H_4P_08$ is observed up to 20% strain. After that, cracks start to appear in the pillar joining the vertical section, which occurs only in the upper and middle horizontal sections. The lower horizontal sections have no significant changes, and they remain intact. The initial deformation behavior of $H_2P_18$ is similar to $H_2P_08$ (layer-by-layer deformations) up to 10% strain, and after that, the deformation behavior is similar to $H_4P_08$ (separation of vertical section). The pores between horizontal sections collapse for both structures, $H_3P_18$ and $H_2P_38$. All horizontal sections become deformed up to 40% strain.

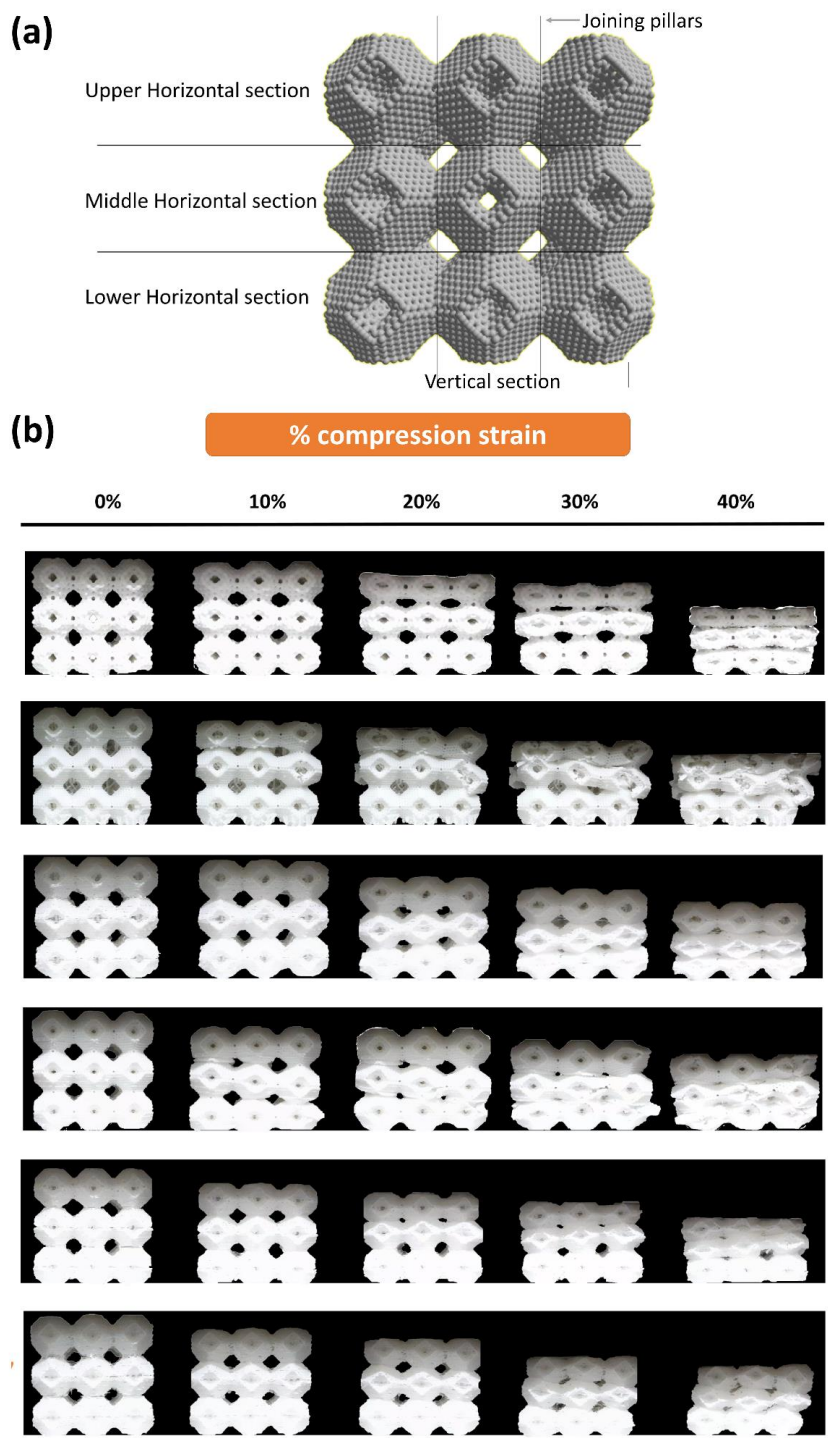

**Figure 6.** AIPS deformation behavior under experimental compression tests. (a) Schematic of the sections of a typical AIPS structure. (b) Experimental photo snapshots at 0, 10, 20, 30, and 40% strain.

In order to better understand the deformation behavior observed in the experimental snapshot images, we have also taken MD snapshots of the deformed atomic models at different strains, such as 0, 10, 20, 30, and 40% strains, as shown in **Fig. 7**. The von Mises's stress variation along the structure is mapped with different colors (blue and red for low and high-stress values, respectively). The stress distribution in all the structures is dependent on their topology and loading direction. At the 10% strain snapshot, we can see a trend of decreasing stress accumulation among two-layered structures ($H_2P_08$, $H_2P_18$, and $H_2P_38$). Based on color mapping, we can conclude that $H_2P_18$ can stand a higher load compared to $H_2P_38$, which can explain better specific stress properties. Simulated deformation behavior of all the AIPS structures are provided in the Supplementary Videos **SV2**.

On the other hand, the uniform stress distribution in $H_2P_38$ contributes to absorbing more energy, as compared to $H_2P_18$. In a 20% strain condition, we observed a large deformation in the top layer, as compared to the bottom one, which can explain the flat plateau region in the stress-strain curve for all AIPS. The above theoretical observation is consistent with the experimental photo snapshot shown in **Fig. 6**. Further deformations, i.e., 30 and 40%, show densification of the layers, which is consistent again with the experimental observations. In the cases of three-layered AIPS, we observed a similar behavior to the two-layered ones. The four-layered AIPS ($H_4P_08$) shows a similar deformation behavior to that of $H_3P_18$, which is consistent with stress-strain curves and experimental photo snapshots. The stress accumulation is higher along the loading direction (red color). The stress distribution is more homogeneous (blue and white color) perpendicular to the loading direction. The higher stress accumulation along the joining pillars induces structural deformations in those regions, which is consistent in our experimental snapshots as well. In order to better understand the role of individual layers, we present the stress distribution of individual layers for 2, 3, and 4 layers AIPS in **Figs. 8-10**.

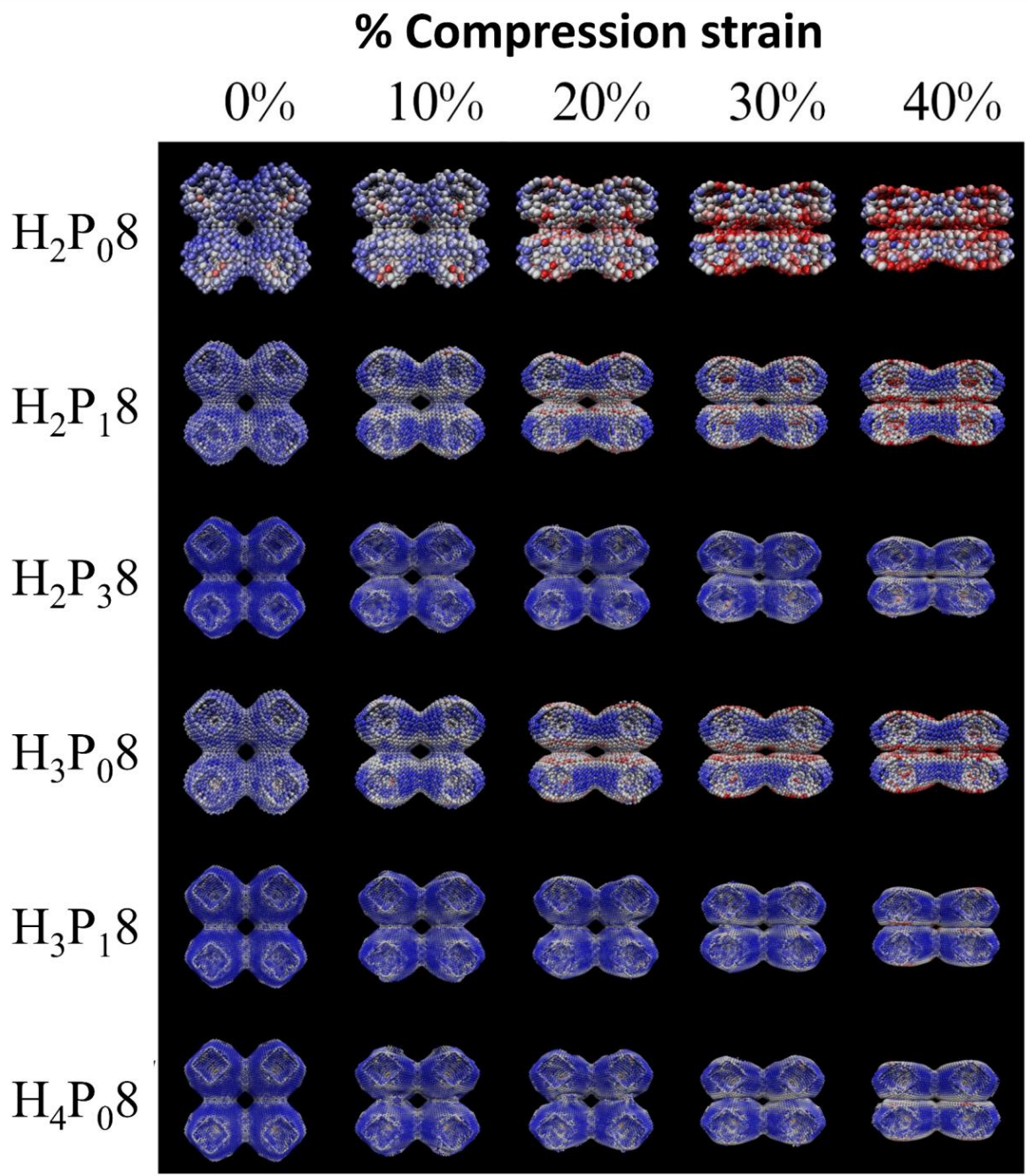

**Figure 7.** MD snapshots for different AIPS at 0, 10, 20, 30, and 40% strain. The color gradient scale bar is shown along the compression direction.

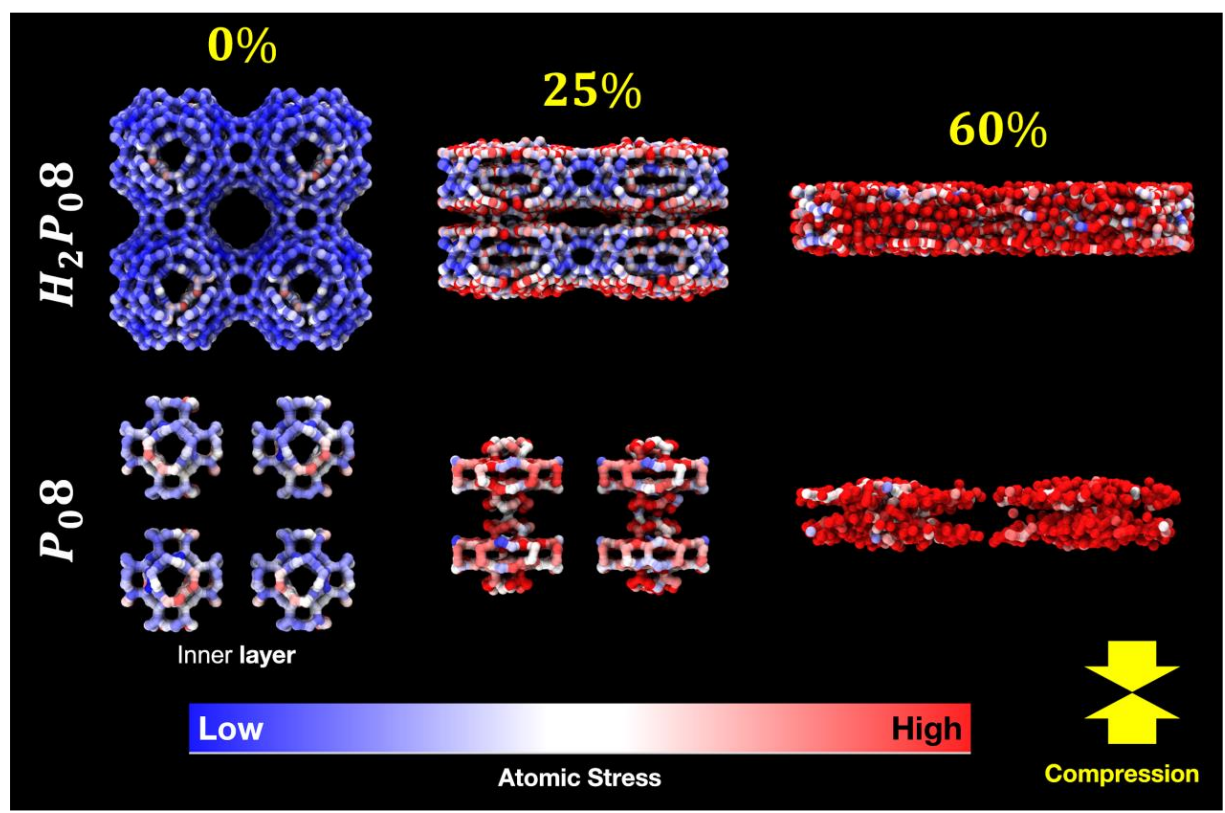

**Figure 8.** MD snapshots for different layers of $H_2P_08$, at 0, 25, and 60% strain. The color gradient scale bar is shown along the compression direction.

In the case of $H_2P_08$, consisting of two interlocked layers (inner $P_08$ and outer schwarzite), as shown in **Fig. 8**, the stress after 25% strain shows a high-stress accumulation in the inner layer (P08), as compared to the outer layer. The outer layer stress accumulation depends on the loading direction. The interconnect region of the building blocks exhibits the highest stress accumulation, as it has a low loading area, which is consistent with the experimental observation in **Fig. 6**. The radial direction or surfaces show higher stress accumulation due to the curvature effects. The inner layer P08 shows high-stress accumulation overall. Based on the stress map, we can conclude that the inner layer takes the load and provides high strength to the $H_2P_08$.

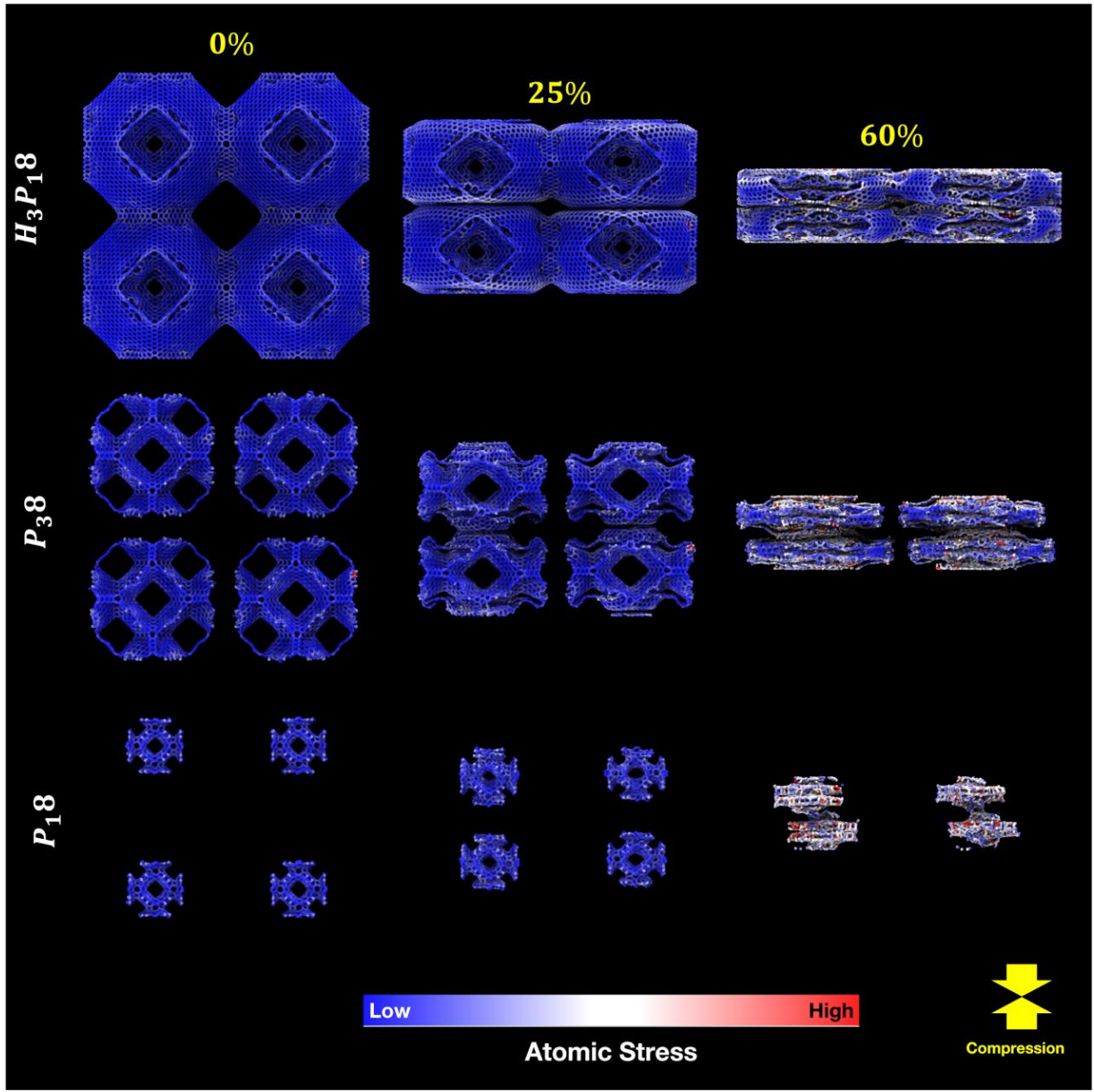

**Figure 9.** MD snapshots for different layers of $H_3P_18$ are 0, 25, and 60% strain. The color gradient scale bar is shown along the compression direction.

In the case of $H_3P_18$, consisting of three interlocked layers ($P_18$, $P_38$, and outer schwarzite), as shown in **Fig. 9**, the stress accumulation after 25% strain shows high stress in the inner layer ($P_18$) as compared to $P_38$ and the outer layer. Similar to $H_2P_08$, the interconnect region of the building blocks exhibits the highest stress accumulation, as it has a low loading area. The middle layer P38 is compressed under loading and improves energy absorption. Based on the stress map, we can conclude that the inner layer takes the load and provides high strength to the $H_2P_08$. Overall, the three-layered $H_3P_18$ exhibits less stress accumulation, which means it can absorb more mechanical energy, which is consistent with our experimental observations. The four-layered $H_4P_08$ shows similar stress distribution as $H_3P_18$, except that an additional layer in the structures reduces the stress accumulations in adjacent layers (as shown in supplementary **Fig. S1**). Also, the additional core layer takes/absorbs energy and improves the mechanical energy absorption, as observed in experiments and theory.

## 4. Conclusion

We used modified porous schwarzites' unit cells to create a new class of hybrid porous structure, which we named architectured interlocked Petal-Schwarzites (AIPS). In order to investigate the role of layers and topology on the AIPS mechanical behavior, we have chosen three different AIPS classes consisting of two, three, and four layers. In each of these classes, we built a combination of different building blocks, which can be used to tune porosity and mechanical response. Here, we report a total of six different AIPS. The atomic models were translated into structures that were 3D-printed. Similar to pure schwarzites, the atomic (MD simulations) and 3D-printed hybrid AIPS structures show similar trends in mechanical behavior (under compression), showing that the topology dominates the scale-independent mechanical behavior. The individual layer stress concentration mapping reveals the role of the interlocked layers. As we increase the number of interlocked layers and topology, we can tune the mechanical response (yield strength, modulus, and specific energy absorption properties). The present work opens up new perspectives to create novel high-strength architectures with enhanced energy absorption properties.

**References**

[1]     Z. Xin, X. Zhang, Y. Duan, W. Xu, Nacre-inspired design of CFRP composite for improved energy absorption properties, Compos. Struct. 184 (2018) 102–109. https://doi.org/10.1016/j.compstruct.2017.09.075.

[2]     L. Zorzetto, D. Ruffoni, Wood-Inspired 3D-Printed Helical Composites with Tunable and Enhanced Mechanical Performance, Adv. Funct. Mater. 29 (2019) 1805888. https://doi.org/10.1002/adfm.201805888.

[3]     N.S. Ha, G. Lu, A review of recent research on bio-inspired structures and materials for energy absorption applications, Compos. Part B Eng. 181 (2020) 107496. https://doi.org/10.1016/j.compositesb.2019.107496.

[4]     Y. Xiao, H. Yin, H. Fang, G. Wen, Crashworthiness design of horsetail-bionic thin-walled structures under axial dynamic loading, Int. J. Mech. Mater. Des. 12 (2016) 563–576. https://doi.org/10.1007/s10999-016-9341-6.

[5]     Y. Chen, T. Li, Z. Jia, F. Scarpa, C.W. Yao, L. Wang, 3D printed hierarchical honeycombs with shape integrity under large compressive deformations, Mater. Des. 137 (2018) 226–234. https://doi.org/10.1016/j.matdes.2017.10.028.

[6]     H. Rhee, M.T. Tucker, W.R. Whittington, M.F. Horstemeyer, H. Lim, Structure-property responses of bio-inspired synthetic foams at low and high strain rates, Sci. Eng. Compos. Mater. 22 (2015). https://doi.org/10.1515/secm-2013-0238.

[7]     Z. Tang, N.A. Kotov, S. Magonov, B. Ozturk, Nanostructured artificial nacre, Nat. Mater. 2 (2003) 413–418. https://doi.org/10.1038/nmat906.

[8]     S.M. Sajadi, P.S. Owuor, S. Schara, C.F. Woellner, V. Rodrigues, R. Vajtai, J. Lou, D.S. Galvão, C.S. Tiwary, P.M. Ajayan, Multiscale Geometric Design Principles Applied to 3D

Printed Schwarzites, Adv. Mater. 30 (2018) 1704820. https://doi.org/10.1002/adma.201704820.

[9] S.M. Sajadi, C.F. Woellner, P. Ramesh, S.L. Eichmann, Q. Sun, P.J. Boul, C.J. Thaemlitz, M.M. Rahman, R.H. Baughman, D.S. Galvão, C.S. Tiwary, P.M. Ajayan, 3D Printed Tubulanes as Lightweight Hypervelocity Impact Resistant Structures, Small. 15 (2019) 1904747. https://doi.org/10.1002/smll.201904747.

[10] R.S. Ambekar, E.F. Oliveira, B. Kushwaha, V. Pal, L.D. Machado, S.M. Sajadi, R.H. Baughman, P.M. Ajayan, A.K. Roy, D.S. Galvao, C.S. Tiwary, On the mechanical properties of atomic and 3D printed zeolite-templated carbon nanotube networks, Addit. Manuf. 37 (2021) 101628. https://doi.org/10.1016/j.addma.2020.101628.

[11] R.S. Ambekar, B. Kushwaha, P. Sharma, F. Bosia, M. Fraldi, N.M. Pugno, C.S. Tiwary, Topologically engineered 3D printed architectures with superior mechanical strength, Mater. Today. 48 (2021) 72–94. https://doi.org/10.1016/j.mattod.2021.03.014.

[12] R.S. Ambekar, E.F. Oliveira, B. Kushwaha, V. Pal, P.M. Ajayan, A.K. Roy, D.S. Galvao, C.S. Tiwary, Flexure resistant 3D printed zeolite-inspired structures, Addit. Manuf. 47 (2021) 102297. https://doi.org/10.1016/j.addma.2021.102297.

[13] H. Terrones, M. Terrones, F. López–Urías, J.A. Rodríguez–Manzo, A.L. Mackay, Shape and complexity at the atomic scale: the case of layered nanomaterials, Philos. Trans. R. Soc. London. Ser. A Math. Phys. Eng. Sci. 362 (2004) 2039–2063. https://doi.org/10.1098/rsta.2004.1440.

[14] L.C. Felix, C.F. Woellner, D.S. Galvao, Mechanical and energy-absorption properties of schwarzites, Carbon N. Y. 157 (2020) 670–680. https://doi.org/10.1016/j.carbon.2019.10.066.

[15] R.S. Ambekar, I. Mohanty, S. Kishore, R. Das, V. Pal, B. Kushwaha, A.K. Roy, S. Kumar Kar, C.S. Tiwary, Atomic Scale Structure Inspired 3D-Printed Porous Structures with Tunable Mechanical Response, Adv. Eng. Mater. (2021) 2001428. https://doi.org/10.1002/adem.202001428.

[16] S.M. Sajadi, P.S. Owuor, R. Vajtai, J. Lou, R.S. Ayyagari, C.S. Tiwary, P.M. Ajayan, Boxception: Impact Resistance Structure Using 3D Printing, Adv. Eng. Mater. 21 (2019) 1900167. https://doi.org/10.1002/adem.201900167.

[17] S.M. Sajadi, L. Vásárhelyi, R. Mousavi, A.H. Rahmati, Z. Kónya, Á. Kukovecz, T. Arif, T. Filleter, R. Vajtai, P. Boul, Z. Pang, T. Li, C.S. Tiwary, M.M. Rahman, P.M. Ajayan, Damage-tolerant 3D-printed ceramics via conformal coating, Sci. Adv. 7 (2021). https://doi.org/10.1126/sciadv.abc5028.

[18] T. Li, Y. Chen, X. Hu, Y. Li, L. Wang, Exploiting negative Poisson's ratio to design 3D-printed composites with enhanced mechanical properties, Mater. Des. 142 (2018) 247–258. https://doi.org/10.1016/j.matdes.2018.01.034.

[19] Y. Chen, Z. Yu, Y. Ye, Y. Zhang, G. Li, F. Jiang, Superelastic, Hygroscopic, and Ionic Conducting Cellulose Nanofibril Monoliths by 3D Printing, ACS Nano. 15 (2021) 1869–1879. https://doi.org/10.1021/acsnano.0c10577.

[20] S. Kee, M.A. Haque, D. Corzo, H.N. Alshareef, D. Baran, Self-Healing and Stretchable 3D-Printed Organic Thermoelectrics, Adv. Funct. Mater. 29 (2019) 1905426. https://doi.org/10.1002/adfm.201905426.